\documentclass[floatfix,aps,prl,preprint, superscriptaddress]{revtex4-2}

\usepackage{upgreek}
\usepackage{graphicx}
\usepackage[]{epsfig}
\usepackage{times,amsmath,amssymb}

\usepackage{color}
\usepackage[colorlinks = true,
            linkcolor = blue,
            urlcolor  = blue,
            citecolor = blue,
            anchorcolor = blue]{hyperref}

\begin{document}

\title{Microwave dependent quantum transport characteristics in GaN/AlGaN FETs}

\author{Motoya Shinozaki}
\email[These authors contributed equally to this work]{}
\affiliation{WPI Advanced Institute for Materials Research, Tohoku University, 2-1-1 Katahira, Aoba-ku, Sendai 980-8577, Japan}

\author{Takaya Abe}
\email[These authors contributed equally to this work]{}
\affiliation{Research Institute of Electrical Communication, Tohoku University, 2-1-1 Katahira, Aoba-ku, Sendai 980-8577, Japan}
\affiliation{Department of Electronic Engineering, Graduate School of Engineering, Tohoku University, Aoba 6-6-05, Aramaki, Aoba-Ku, Sendai 980-8579, Japan}

\author{Kazuma Matsumura}
\affiliation{Research Institute of Electrical Communication, Tohoku University, 2-1-1 Katahira, Aoba-ku, Sendai 980-8577, Japan}
\affiliation{Department of Electronic Engineering, Graduate School of Engineering, Tohoku University, Aoba 6-6-05, Aramaki, Aoba-Ku, Sendai 980-8579, Japan}

\author{Takumi Aizawa}
\affiliation{Research Institute of Electrical Communication, Tohoku University, 2-1-1 Katahira, Aoba-ku, Sendai 980-8577, Japan}
\affiliation{Department of Electronic Engineering, Graduate School of Engineering, Tohoku University, Aoba 6-6-05, Aramaki, Aoba-Ku, Sendai 980-8579, Japan}

\author{Takashi Kumasaka}
\affiliation{Research Institute of Electrical Communication, Tohoku University, 2-1-1 Katahira, Aoba-ku, Sendai 980-8577, Japan}

\author{Tomohiro Otsuka}
\email[]{tomohiro.otsuka@tohoku.ac.jp}
\affiliation{WPI Advanced Institute for Materials Research, Tohoku University, 2-1-1 Katahira, Aoba-ku, Sendai 980-8577, Japan}
\affiliation{Research Institute of Electrical Communication, Tohoku University, 2-1-1 Katahira, Aoba-ku, Sendai 980-8577, Japan}
\affiliation{Department of Electronic Engineering, Graduate School of Engineering, Tohoku University, Aoba 6-6-05, Aramaki, Aoba-Ku, Sendai 980-8579, Japan}
\affiliation{Center for Science and Innovation in Spintronics, Tohoku University, 2-1-1 Katahira, Aoba-ku, Sendai 980-8577, Japan}
\affiliation{Center for Emergent Matter Science, RIKEN, 2-1 Hirosawa, Wako, Saitama 351-0198, Japan}

\begin{abstract}
Defects in semiconductors, traditionally seen as detrimental to electronic device performance, have emerged as potential assets in quantum technologies due to their unique quantum properties. 
This study investigates the interaction between defects and quantum electron transport in GaN/AlGaN field-effect transistors, highlighting the observation of Fano resonances at low temperatures. 
We observe the resonance spectra and their dependence on gate voltage and magnetic fields.
To explain the observed behavior, we construct the possible scenario as a Fano interferometer with finite width.
Our findings reveal the potential of semiconductor defects to contribute to the development of quantum information processing, providing their role to key components in next-generation quantum devices.
\end{abstract}

\maketitle

In electronics applications, defects in solid states are typically viewed as detrimental, impacting device quality, integrated circuit yield, and electrical noise~\cite{chang1994flicker}. 
Conversely, from a quantum mechanical perspective, the isolated energy levels introduced by these defects have garnered significant interest~\cite{wolfowicz2021quantum}. 
Many studies have explored the unique properties of defects for their potential applications in advanced optics~\cite{aharonovich2011diamond} and quantum computation~\cite{weber2010quantum, koehl2011room, wolfowicz2013atomic, widmann2015coherent}. 
These isolated energy levels can act as quantum bits (qubits) for information processing, a high-sensitive magnetic sensor, and emissive centers in quantum optics, offering a pathway to exploit defects for innovative technologies~\cite{anderson2019electrical}. 

Quantum transport phenomena in solid states, such as quantum Hall effect, single-electron transport, and Josephson tunneling are quite important for quantum devices, which are based on quantum mechanics and are expected to beyond conventional electronic devices~\cite{Klitzing1980, Jeckelmann2003, Kastner1992, Chen1996, Makhlin2001}. 
Among them, those observed in semiconductor nano-structures are attractive from the viewpoint of compatibility with semiconductor manufacturing technologies~\cite{Tarucha1996, Hanson2007, Li2006}. 
We can design the various nano-structures using the technologies and control properties of the quantum transport phenomena in those. Quantum dots~\cite{Gaudreau2006prl, braakman2013long, Takakura2014, Ito2016, Ito2018, Noiri2018} are one of the prospective structures for the realization of quantum computation~\cite{Loss1998}. 

Exploring the interaction between defects and quantum electron transport in semiconductor devices presents a fascinating research avenue. 
This interplay is crucial for understanding how defects influence quantum behavior in semiconductors, potentially providing interesting functionalities in quantum sensing and information processing technologies. 
By investigating these interactions, we can not only gain deeper insights into fundamental physics but also improve the design and performance of semiconductor devices.
Experiments previously reported~\cite{ferrus2009cryogenic, rossi2010microwave, erfani2011microwave, Tenorio-Pearl2017} have turned the spotlight on defects or trapped states at the silicon-based semiconductor/oxide interface, traditionally considered a source of electrical noise in field-effect transistors (FETs). 
Moreover, relatively long coherence times of 1 to 40 $\upmu \mathrm{s}$ have been demonstrated by controlling the defects as a quantum two-level system~\cite{Tenorio-Pearl2017}. 
These findings underscore the potential of such defects not merely as nuisances but as gateways to new qubits and physical insights in semiconductor devices.

In this study, we investigate the microwave-dependent quantum transport in the low-temperature conduction of GaN/AlGaN FETs~\cite{ambacher1999two, manfra2004electron, thillosen2006weak, shchepetilnikov2018electron}.
Previously, we observed the formation of quantum dots in GaN/AlGaN FETs, near the pinch-off condition at a cryogenic temperature~\cite{Otsuka2020, matsumura2023channel, fujiwara2023wide}. 
The possible mechanism is the disturbed potential caused by impurities and defects near the conduction channel of the FETs. 
Because the transport is modified by the states of the impurities and defects, it can be able to observe resonance spectra by applying microwaves, as in previous studies~\cite{ferrus2009cryogenic, rossi2010microwave, erfani2011microwave, Tenorio-Pearl2017}.
We also measure the gate voltage and in-plane magnetic field dependence of the microwave resonance to discuss the possible current path.

\begin{figure}
\begin{center}
  \includegraphics{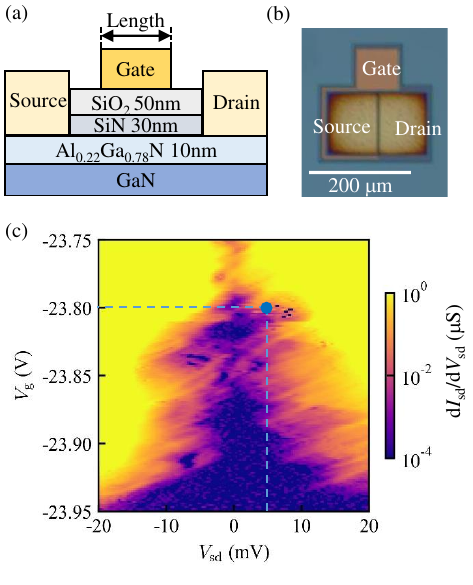}
  \caption{(a) Schematic of the layer structure of the device.
  (b) Optical image of the device.
  (c) A typical result of electron transport measurements near the pinch-off voltage of the FET at 2.3 K. Dashed lines indicate a operation point of microwave resonance measurement for Fig.~\ref{resonances} (b). 
  }
  \label{device}
\end{center}
\end{figure}

Figure~\ref{device}(a) presents a diagram of the device's layered structure. Layers of GaN and AlGaN are deposited on a silicon substrate through chemical vapor deposition. 
A two-dimensional electron gas forms at the GaN and AlGaN interface, exhibiting typical electron densities and mobilities of $6.7\times10^{12}\; \mathrm{cm}^{-2}$ and $1670 \; \mathrm{cm}^{2}\mathrm{V}^{-1}\mathrm{s}^{-1}$, respectively. 
Titanium and aluminum are employed for the source and drain contacts. 
The gate insulators are made of SiN and $\mathrm{SiO}_2$, with a TiN gate electrode layered above.
The FET has gate dimensions of 0.6 $\upmu \mathrm{m}$ in length and 150 $\upmu \mathrm{m}$ in width.
The optical image of the device is shown in Fig.~\ref{device}(b). 
The source-drain current $I_{\mathrm{sd}}$ is measured as a function of the source-drain voltage $V_{\mathrm{sd}}$ and the gate voltage $V_\mathrm{g}$. 
A microwave from a signal generator is applied to the gate electrode through a homemade cryogenic bias tee.
To measure the electron transport properties at low temperatures of 2.3 K, we use a helium depressurization refrigerator. 

Figure~\ref{device}(c) shows a typical electron transport characteristics near the pinch-off state of the FET at 2.3 K. 
The differential conductance $\mathrm{d}I_{\mathrm{sd}}/{\mathrm{d}V_{\mathrm{sd}}}$ are measured as a function of $V_{\mathrm{sd}}$ and $V_\mathrm{g}$. 
We observed overlapped Coulomb diamond structures because of the formation of multiple quantum dots in the conduction channel. 
This is due to the potential fluctuations by impurities or defects near the conduction channel~\cite{Otsuka2020, matsumura2023channel}. 
The blue point in Fig.~\ref{device}(c) shows the typical measurement point for the microwave resonance measurement. 
We set $V_{\mathrm{sd}}$ and $V_\mathrm{g}$ at this point and measure $I_{\mathrm{sd}}$ with sweeping microwave frequency.

\begin{figure}
\begin{center}
  \includegraphics{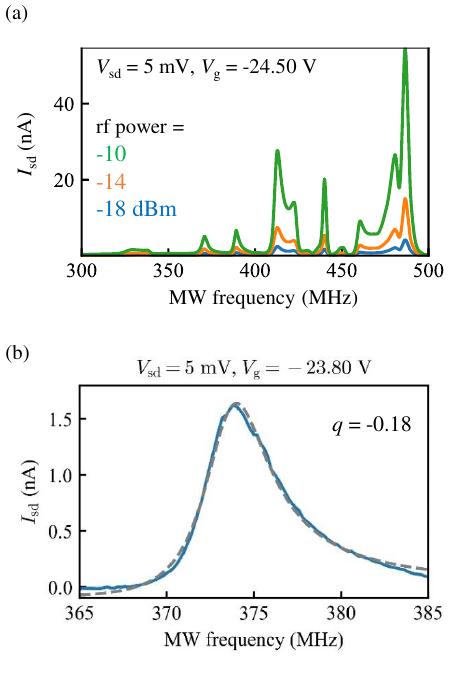}
  \caption{(a) Dependence of the resonance spectra on microwave power. (b) A detailed spectrum around 375 MHz. Solid and dashed lines correspond to the experimental and fitting results, respectively.
  }
  \label{resonances}
\end{center}
\end{figure}

Figure~\ref{resonances}(a) presents a typical result of the microwave resonance experiments and its power dependence.
When sweeping the microwave frequency, the $I_{\mathrm{sd}}$ is modulated, revealing several resonance spectra. 
The observed resonance is attributed to opening defect levels as current paths upon the application of microwaves.
This phenomenon aligns with findings from previous studies~\cite{Tenorio-Pearl2017} and is identified as a Fano resonance. 
Fano resonance is characterized by a unique interference effect between a discrete quantum state and a continuum, leading to an asymmetric line shape in the resonance spectra~\cite{Fano1961}.
The line shape of the resonance spectra remains unchanged across different microwave power levels. 
Notably, the resonance spectra exhibit a linewidth within the range of 1-10 MHz. 
In contrast, a previous study involving Si FETs reported a linewidth of approximately 60 kHz~\cite{erfani2011microwave, Tenorio-Pearl2017}. 
The observed broader linewidths in our study may be attributed to the shorter lifetimes and broadening of the discrete energy levels contributing to resonance in GaN FETs.
Note that these resonance peaks are reproducible during the sample is maintained at low temperatures.
We also perform the measurements after subjecting the sample to thermal cycling and light exposure. 
These processes altered the detailed structures of the resonance peaks, although the overall frequency range of the peaks remained consistent. 
The observed changes are likely due to the redistribution of charged impurities and defects induced by thermal and optical cycles.
These resonance spectra are not observed in the pinch-off state ($V_{\mathrm{g}} = -30\ \mathrm{V}$). 
Additionally, we measure the transmission coefficient $S_{21}$ of the high-frequency circuit line using a network analyzer and confirm that there are no reflection peaks corresponding to the frequency at which the resonance spectra are observed. 

Here, we focus on the resonance spectra around 375 MHz as shown in Fig.~\ref{resonances}(b).
This resonance spectra with the Fano-like line shape are described by the Fano formula~\cite{Fano1961, Kobayashi2002, Kobayashi2003, Otsuka2007, Tenorio-Pearl2017}:
\begin{equation}
F(E) = \frac{(E + q)^2}{E^2 + 1},
\label{fano}
\end{equation}
where $F$ is the transmission probability and $E$ the normalized energy expressed by $(f-f_0)/(\delta f/2)$. 
$f_0$ and $\delta f$ are the resonance frequency and the linewidth, respectively. 
$q$ is the Fano parameter representing the resonance peak's asymmetry. 
The results of the fitting using this equation with a free parameter $q$ are shown in Fig.~\ref{resonances}(b), and well reproduced the experimental data.

\begin{figure}
\begin{center}
  \includegraphics{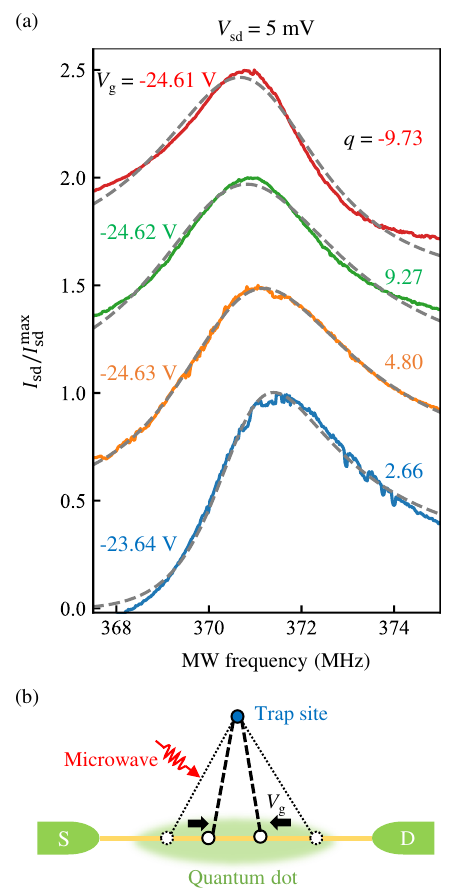}
  \caption{(a) Gate bias dependence of the resonance spectrum. (b) The model of Fano effect with finite width. The coupling width is changed by applying $V_{\mathrm{g}}$.
  }
  \label{bias}
\end{center}
\end{figure}

Fig.~\ref{bias}(a) shows the $V_\mathrm{g}$ dependence of the microwave resonance spectrum. 
In this measurement, $V_{\mathrm{sd}}$ is also fixed at 5 mV, and $V_{\mathrm{g}}$ is measured at four points between -23.64 V and -24.61 V in 0.01 V increments. 
For comparison, the measurement results were normalized to the maximum value of the resonance spectrum, $I_{\mathrm{sd}}^{\mathrm{max}}$, and an offset is added. 
At a glance, the shape of the resonance spectrum is modulated by the change of $V_{\mathrm{g}}$.
We fit the spectrum by using Eq.~\ref{fano}, and the extracted parameters $q$ are shown in Fig.~\ref{bias}(a). 
The value and sign of $q$, which represents the asymmetry of the peak, changes by changing the $V_{\mathrm{g}}$.
There is another resonance spectrum around 97.5 MHz obtaining similar $V_\mathrm{g}$ dependence, which is detailed in Appendix A.

To explain the change of the line shape by the change of  $V_{\mathrm{g}}$, we construct a possible model.
It is a Fano interferometer with finite width as shown in Fig.~\ref{bias}(b)~\cite{Otsuka2007}.
The trapping sites in the FET, caused by impurities or defects, exhibit an energy level distribution influenced by temperature and/or carrier lifetime.
The width of the Coulomb peak also indicates the energy level distribution within the system, primarily determined by the tunneling rate between the quantum dots and the electrodes.
In GaN/AlGaN quantum dots, the typical energy distribution spans a few millielectronvolts~\cite{matsumura2023channel}, a range that is considerably broader than the linewidth observed in the Fano resonance spectrum. 
This suggests that the trapping sites within our device act as effective discrete states, influencing the microwave resonance.
The trapping site couples to the current path with finite width: several points in the current path coupled to the trapping site.
With the change of $V_{\mathrm{g}}$, some parameters of the system, such as coupling between continuous and discrete paths, are modified.
This induces the change in the interference and we observe the change of the Fano resonance spectrum.
We perform a tight-binding calculation to support our scenario in Appendix B.

\begin{figure}
\begin{center}
  \includegraphics{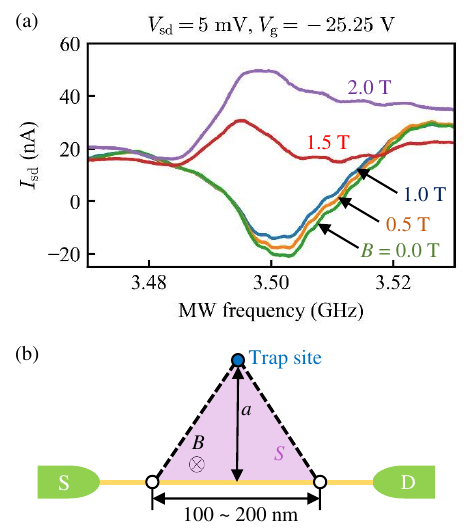}
  \caption{(a) In-plane magnetic field dependence of the resonance spectrum. (b) The assumed model with a finite width to create a region that crosses the magnetic flux quantum.
  }
  \label{magnetic}
\end{center}
\end{figure}

Several resonance spectra show a change in the line shape by applying the in-plane magnetic field to the two-dimensional electron gas, as shown in Fig.~\ref{magnetic}.
In this measurement, the bias was fixed at $V_{\mathrm{sd}} = 5\ \mathrm{mV}$ and $V_{\mathrm{g}} = -25.25\ \mathrm{V}$.
The in-plane magnetic field $B$ changes from 0 T to 2 T in 0.5 T steps. 
Around 1.5 T, a steep change in the shape of the resonance spectrum is observed. 
A similar trend was observed in the reverse magnetic field direction.
As to the possible mechanism, the additional phase induced by the magnetic field between the interference paths can be expected~\cite{Kobayashi2002, Kobayashi2003}.
Note that the negative current is observed in some cases, and the abrupt change occurs around 1.5~T.
Only an additional phase cannot explain these behaviors.
One of the possible scenarios of additional mechanisms is the photon-assisted back currents~\cite{KouwenhovenPRL1994}.

For the discussion, it is assumed that the abrupt change observed around 1.5 T is due to phase inversion, with a magnetic flux quantum $\Phi_0$ entering the triangular region $S$ depicted in Figure~\ref{magnetic}(b).
Taking into account that the coupling width ranges from 100 to 200 nm, which corresponds to the potential fluctuation in the conduction channel as mentioned in previous work~\cite{matsumura2023channel}, we estimate the distance between the current path and the trapping site to be $a = 14\sim28$ nm from the $\Phi_0=BS$.
This value aligns with the geometry of our device, supporting the hypothesis that the resonance spectrum is induced by defects or impurities near the conduction channel~\cite{guerra2010comparison}.
Furthermore, there is no response on the resonance spectrum when the magnetic field is applied in the 2DEG normal direction, which indicates that the area pierced by magnetic flux is not formed in the 2DEG plane.

In conclusion, we measure electron transport under microwave irradiation at low-temperature in GaN/AlGaN FETs.
Asymmetric resonance spectra are observed near the pinch-off of the FET channel, where the quantum dots form.
The shape of some peaks is modified by the gate voltage and the in-plane magnetic field. 
These behaviors are explained by the Fano interferometer.
From this model, we estimate that the Fano resonances are induced due to interaction between the conduction channel and defect sites in AlGaN or insulating layers. 
Our findings open avenues for employing GaN/AlGaN FETs in quantum devices, offering potential applications in defect-based qubit manipulation and evaluating defects in semiconductor devices. 

\section{Acknowledgements}
The authors thank N, Ito, T, Tanaka, K, Nakahara, Y. Tokura, 
RIEC Fundamental Technology Center and the Laboratory for Nanoelectronics and Spintronics 
for fruitful discussions and technical supports.
Part of this work is supported by  
Rohm Collaboration Project, 
MEXT Leading Initiative for Excellent Young Researchers, 
Grants-in-Aid for Scientific Research (21K18592, 23H01789, 23H04490),
Hattori Hokokai Foudation Research Grant, 
The Foundation for Technology Promotion of Electronic Circuit Board, 
Iketani Science and Technology Foundation Research Grant, 
and FRiD Tohoku University. 

\section{Appendix A: Gate voltage dependence of the resonance spectrum}
\begin{figure}
\begin{center}
  \includegraphics{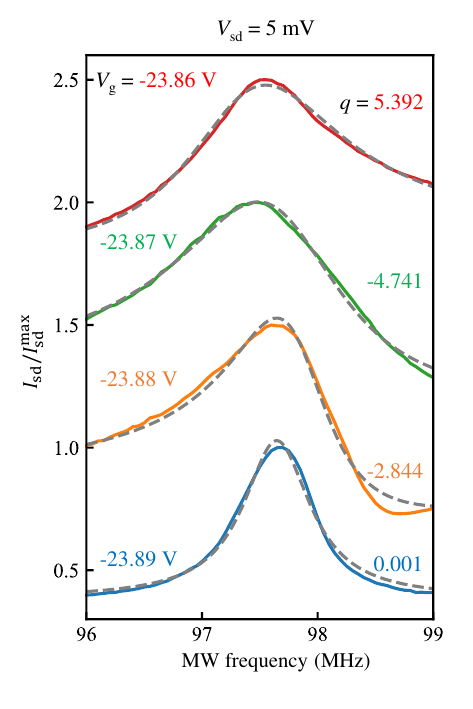}
  \caption{Gate voltage dependence of the resonance spectrum around 97.5 MHz.}
  \label{fig_S1}
\end{center}
\end{figure}

Here, we present another resonance spectrum influenced by the gate voltage in Fig.~\ref{fig_S1}. 
This behavior is similar to that we observed with the resonance spectrum approximately at 371 MHz, as illustrated in Fig.~\ref{bias}.

\section{Appendix B: Tight binding calculation}
\begin{figure}
\begin{center}
  \includegraphics{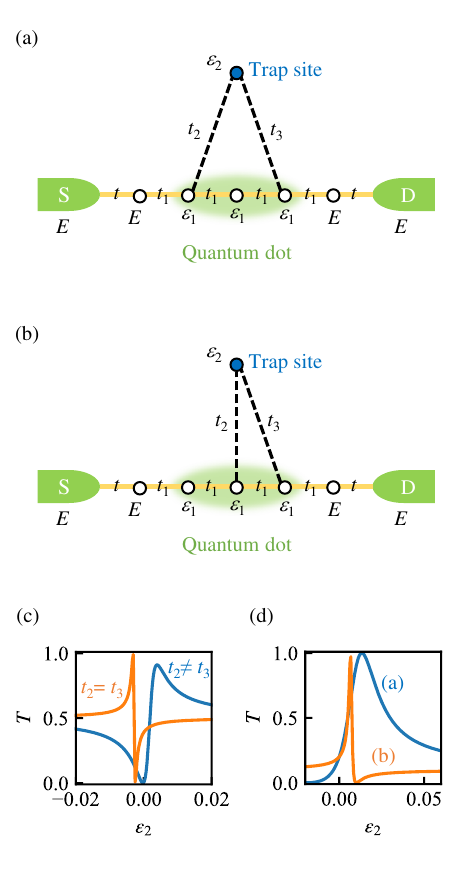}
  \caption{(a, b) A tight-binding model representation of our device system. It is hypothesized that the gate voltage modulates the coupling of the site to the trap site. (c, d) The calculated transmission probabilities for the assumed models.}
  \label{fig_S2}
\end{center}
\end{figure}

We perform a tight binding calculation to support our Fano effect model. 
We assume a one-dimensional model considering dots site energy and nearest neighbor hopping, where one trap site is coupled as shown in Figs.~\ref{fig_S2} (a) and (b).
Here, $\varepsilon_1$, $\varepsilon_2$, $t$, $t_{1}$, $t_{2}$, and $t_{3}$ are the channel dots energy, the trap site energy, and hopping parameters, respectively. 
The retarded Green's function $\mathbf{G}^R$ is described as
\begin{equation}
\mathbf{G}^R = (E\mathbf{I}-\mathbf{H}-\mathbf{\Sigma}^R_L-\mathbf{\Sigma}^R_R)^{-1},
\end{equation}
where $E$, $\mathbf{I}$, $\mathbf{H}$, and $\mathbf{\Sigma}^R_{L, R}$ are the Fermi energy of left and right lead electrodes, the identity matrix, the Hamiltonian of the conductor, and the self-energy of our model, respectively.
From the Meir-Wingrenn fomula~\cite{meir1992landauer}, the transmission probability can be derived as
\begin{equation}
T = \mathrm{Tr}[\mathbf{\Gamma}_R \mathbf{G}^R \mathbf{\Gamma}_L \mathbf{G}^{R\dagger}],
\label{eq3}
\end{equation}
\begin{equation}
\mathbf{\Gamma}_{R, L} = -2\mathrm{Im}(\mathbf{\Sigma}^R_{L, R}).
\end{equation}
Here, we calculate the transmission probability $T$ by Eq.~\ref{eq3} under two models as shown in Figs.~\ref{fig_S2}(a) and (b).
Figure~\ref{fig_S2}(c) presents a comparison between conditions $t_{2}=t_{3}$ and $t_{2} \neq t_{3}$ in the model of Fig.~\ref{fig_S2}(a).
We also present the results obtained by varying the coupling site under the condition $t_{2}=t_{3}$ in Fig.~\ref{fig_S2}(d), corresponding to the models depicted in Figs.~\ref{fig_S2}(a) and (b).
Clearly, the sign of the Fano parameter, resonance frequency (energy), and linewidth are changing.
The changes observed in the spectrum during the experiments may be attributed to the modulation of these and other potential parameters, such as $\varepsilon_{1}$, by the gate voltage.

\bibliography{reference.bib}
\end{document}